# On a new type of solving procedure for Laplace tidal equation.

**Sergey V. Ershkov**

Peoples' Friendship University of Russia (RUDN University),

6 Miklukho-Maklaya Street, Moscow, 117198, Russian Federation,

e-mail: sergej-ershkov@yandex.ru

**Roman V. Shamin**

Moscow Technological University (MIREA),

78 Vernadsky Avenue, Moscow 119454, e-mail: roman@shamin.ru

In this paper, we present a new approach for solving Laplace tidal equations (LTE) which was formulated first in [S.V.Ershkov, "*A Riccati-type solution of Euler-Poisson equations of rigid body rotation over the fixed point*", Acta Mechanica, 228(7), 2719 (2017)] for solving Poisson equations: a new type of the solving procedure for Euler-Poisson equations (rigid body rotation over the fixed point) is implemented here for solving momentum equation of LTE, Laplace tidal equations. Meanwhile, the system of Laplace tidal equations (including continuity equation) has been successfully explored with respect to the existence of analytical way for presentation of the solution.

As the main result, the new ansatz is suggested here for solving LTE: solving momentum equation is reduced to solving system of 3 nonlinear ordinary differential equations of 1-st order in regard to 3 components of the flow velocity (depending on time *t*), along with the continuity equation which determines the spatial part of solution. Nevertheless, the proper elegant *partial* solution has been obtained due to invariant dependence between temporary components of the solution. In addition to this, it is proved here that the system of Laplace tidal equations has not the analytical presentation of solution (in quadratures) in case of nonzero fluid pressure in the Ocean, as well as nonzero total gravitational potential and the centrifugal potential (due to planetary rotation).

1. **Introduction, equations of motion.**

   Laplace tidal equations (LTE, including continuity equation) describe the dynamics of fluid's velocity under the action of potential forces (including gravity) inside the upper layer of Ocean, which is supposed to be located relatively close to the boundary between the Ocean and atmosphere of Earth. It is worth to note that *Pierre-Simon Laplace*, one of the famous scientists in the field of mechanics, have been spending not less than 30 years of his life exclusively first for creating of the self-consistent theory of terrestrial tides in its practical application as the dynamics of tidal waves in ocean basins, and the second for solving the Laplace tidal equations (LTE, named after him). Also, there are a larger amount of previous and recent works concerning analytical treatments with respect to these equations which should be mentioned accordingly [1-6].

   In the co-rotating frame of a Cartesian coordinate system $\vec{r} = \{x, y, z\}$, the linearized (in $\vec{v}$) equations of motion of the ocean that covers the surface of the planet are [7] (*at given initial or boundary conditions*):

$$\nabla \cdot \vec{v} = 0, \tag{1}$$

$$\frac{\partial \vec{v}}{\partial t} + 2\vec{\Omega} \times \vec{v} = -\nabla\left(\frac{p}{\rho} + \Phi + \chi\right), \tag{2}$$

$$-\nabla\left(\frac{p}{\rho} + \Phi + \chi\right) \equiv \begin{Bmatrix} f_x \\ f_y \\ f_z \end{Bmatrix}$$

where $\vec{v}$ is the fluid's velocity, $\vec{v} = \{v_i\}$ ($i = 1, 2, 3$); $\vec{\Omega}$ is the *angular velocity pseudovector* which determines the actual rate (and direction) of Earth's rotation, $\vec{\Omega} = \{\Omega_i\}$, let us note that on a time-scale of few cycles of rotation, each $\Omega_i \cong \Omega_i$ $(x, y, z)$ (not depending on time $t$). Besides, here $\rho$ is the fluid density, $p = p(\vec{r}, t)$ is the pressure, $\Phi(\vec{r})$ is the total gravitational potential, and $\chi(\vec{r})$ is the centrifugal potential (due to planetary rotation); $2\vec{\Omega} \times \vec{v}$ is the Coriolis acceleration (likewise, due to planetary rotation). As for the domain in which the flow occurs and the boundary conditions, let us consider only the Cauchy problem in the whole space.

Furthermore, according to [7], we have neglected the compressibility of the fluid in the Ocean (it means that the aforesaid fluid density $\rho = const$).

## 2. **General presentation of *non-homogeneous* solution for velocity field.**

The momentum Eqn. (2) of LTE is known to be the system of 3 linear partial differential equations, PDEs (in regard to the time *t*) for 3 unknown functions [8-12]: $v_1$, $v_2$, and $v_3$.

Besides, (2) is the system of 3 linear differential equations with all the coefficients depending on time *t*. In accordance with [13] p.71, the general solution of such the aforementioned system should be given as below:

$$\chi_s(t) = \sum_{v=1}^{3} \zeta_{v,s} \cdot \left( \int \left( \frac{\Delta_v}{\Delta} \right) dt + C_v \right), \quad s = 1,2,3, \quad \Delta = \begin{vmatrix} \zeta_{1,1} & \zeta_{1,2} & \zeta_{1,3} \\ \zeta_{2,1} & \zeta_{2,2} & \zeta_{2,3} \\ \zeta_{3,1} & \zeta_{3,2} & \zeta_{3,3} \end{vmatrix} \quad (3)$$

where $\{\chi_s\}$ are the fundamental system of solutions of three Eqns. (2) for components $v_1$, $v_2$, and $v_3$ in regard to the time *t*; $\{\zeta_{v,s}\}$ are the fundamental system of solutions $\vec{v}|_0 = \{U, V, W\}$ for *the corresponding homogeneous* variant of (2), see Eqns. (4) below; $\{C_v\}$ are the set of functions, not depending on time *t*; besides, here we denote as below:

$$\Delta_1 = \begin{vmatrix} f_x & f_y & f_z \\ \zeta_{2,1} & \zeta_{2,2} & \zeta_{2,3} \\ \zeta_{3,1} & \zeta_{3,2} & \zeta_{3,3} \end{vmatrix}, \quad \Delta_2 = \begin{vmatrix} \zeta_{1,1} & \zeta_{1,2} & \zeta_{1,3} \\ f_x & f_y & f_z \\ \zeta_{3,1} & \zeta_{3,2} & \zeta_{3,3} \end{vmatrix}, \quad \Delta_3 = \begin{vmatrix} \zeta_{1,1} & \zeta_{1,2} & \zeta_{1,3} \\ \zeta_{2,1} & \zeta_{2,2} & \zeta_{2,3} \\ f_x & f_y & f_z \end{vmatrix}.$$

It means that the system of Eqns. (2) could be considered as having been solved if we obtain a general solution of *the corresponding homogeneous* system (2), $\vec{v}\,|_0 = \{U, V, W\}$. Let us present in the next section the aforementioned general solution of the corresponding *homogeneous* system, which is presented below (we consider the case $k = 2$ in (2)):

$$\frac{\partial \vec{v}\,|_0}{\partial t} + (k\vec{\Omega}) \times \vec{v}\,|_0 = \vec{0}, \Rightarrow \begin{cases} \dfrac{\partial U}{\partial t} = V \cdot (k\,\Omega_3) - W \cdot (k\,\Omega_2), \\[6pt] \dfrac{\partial V}{\partial t} = W \cdot (k\,\Omega_1) - U \cdot (k\,\Omega_3), \\[6pt] \dfrac{\partial W}{\partial t} = U \cdot (k\,\Omega_2) - V \cdot (k\,\Omega_1)\,. \end{cases} \quad (4)$$

Also, according to the *continuity* equation (1), the appropriate restriction should be valid for identifying of spatial components of velocity field $\vec{v}$ (*non-homogeneous* solution) along with set of functions $\{C_v(x, y, z)\}$ in (3):

$$\frac{\partial v_1}{\partial x} + \frac{\partial v_2}{\partial y} + \frac{\partial v_3}{\partial z} = 0 \qquad (5)$$

which is the PDE-equation of the 1-st kind; $\{v_i\}$ ($i = 1, 2, 3$) depend on variables ($x,y,z, t$).

### 3. **Presentation of the solution of Laplace tidal equations, LTE.**

The system of Eqns. (4) has *the analytical* way to present the general solution [9-12] in regard to the time *t*:

$$U = -\sigma \cdot \left(\frac{2a}{1 + (a^2 + b^2)}\right), \quad V = -\sigma \cdot \left(\frac{2b}{1 + (a^2 + b^2)}\right),$$

$$W = \sigma \cdot \left(\frac{1 - (a^2 + b^2)}{1 + (a^2 + b^2)}\right), \qquad (6)$$

here $\sigma = \sigma(x, y, z)$ is some arbitrary (real) function, given by the initial conditions; the real-valued coefficients $a(x,y,z, t)$, $b(x,y,z, t)$ in (6) are solutions of the mutual system of two *Riccati* ordinary differential equations, ODEs in regard to the time $t$ (here below $k = 2$):

$$\begin{cases} a' = \left(\frac{k \cdot \Omega_2}{2}\right) \cdot a^2 - (k \cdot \Omega_1 \cdot b) \cdot a - \frac{k \cdot \Omega_2}{2}(b^2 - 1) + (k \cdot \Omega_3) \cdot b, \\ b' = -\left(\frac{k \cdot \Omega_1}{2}\right) \cdot b^2 + (k \cdot \Omega_2 \cdot a) \cdot b + \frac{k \cdot \Omega_1}{2} \cdot (a^2 - 1) - (k \cdot \Omega_3) \cdot a. \end{cases} \quad (7)$$

Just to confirm [11] the *Riccati*-type of equations (7): indeed, if we multiply the 1-st of Eqns. (7) on $\Omega_2$, the 2-nd Eq. on $\Omega_1$, then summarize them one to each other properly, we should obtain

$$\Omega_2 \cdot a' - \frac{k}{2}((\Omega_1)^2 + (\Omega_2)^2) \cdot a^2 + k \Omega_1 \cdot \Omega_3 \cdot a + \frac{k(\Omega_1)^2}{2} =$$
$$= -\Omega_1 \cdot b' - \frac{k}{2}((\Omega_1)^2 + (\Omega_2)^2) \cdot b^2 + k \Omega_2 \cdot \Omega_3 \cdot b + \frac{k(\Omega_2)^2}{2}. \quad (8)$$

Equation (8) above is the classical *Riccati* ODE. It describes the evolution of function $a$ in dependence on the function $b$ in regard to the time $t$; such a *Riccati* ODE has no analytical solution in general case [13].

Mathematical procedure for checking of the solution (6)-(7) (which is to be valid for the *homogeneous* LTE equations (4)) has been moved to an **Appendix A1**, with only the resulting formulae left in the main text.

We should especially note that the existence of the solution for LTE equations (1)-(2) is proved to be the question of existence of the proper function $\sigma(x, y, z)$ in (6) and the set of functions $\{C_v (x, y, z)\}$ in (3) of so kind that the PDE-equation (5) should be satisfied under the given initial conditions.

## 4. Solving procedure and the partial solution for equations (7).

System (7) can be reduced as below in case $\Omega_i \cong \Omega_i (x, y, z)$, not depending on time $t$ (we should simply divide one equation onto another; $k = 2$):

$$\frac{da}{db} = \frac{\left(\frac{k \cdot \Omega_2}{2}\right) \cdot a^2 - (k \cdot \Omega_1 \cdot b) \cdot a - \frac{k \cdot \Omega_2}{2}(b^2 - 1) + (k \cdot \Omega_3) \cdot b}{-\left(\frac{k \cdot \Omega_1}{2}\right) \cdot b^2 + (k \cdot \Omega_2 \cdot a) \cdot b + \frac{k \cdot \Omega_1}{2} \cdot (a^2 - 1) - (k \cdot \Omega_3) \cdot a} \qquad (8)$$

Equation (8) is known to be the *Jacobi* equation (see example 1.250 [13]), which has obvious partial solution of a simple kind $a = K \cdot b + L$ ($K = const$, $L = const$) for which are valid the proper conditions below (mathematical procedure of obtaining such the conditions has been moved to an **Appendix A2**, with only the resulting formulae left in the main text):

$$\Omega_2 = -\Omega_1, \quad L = \frac{\Omega_3}{\Omega_1}, \quad K = 1 \qquad (9)$$

Taking into account the demand (9), let us substitute the aforementioned partial solution $a = K \cdot b + L$ into the first of equations (7) (where $k = 2$):

$$a' = \left(-\frac{2 \cdot \Omega_1}{2}\right) \cdot a^2 - 2 \cdot \Omega_1 \cdot (a - \frac{\Omega_3}{\Omega_1}) \cdot a + \frac{2\Omega_1}{2}((a - \frac{\Omega_3}{\Omega_1})^2 - 1) + (2\Omega_3) \cdot (a - \frac{\Omega_3}{\Omega_1}) \Rightarrow$$

$$a' = -2\Omega_1 \cdot a^2 + 2\Omega_3 \cdot a - \left(\Omega_1 + \Omega_3 \cdot \frac{\Omega_3}{\Omega_1}\right)$$

$$\frac{da(t)}{(A \cdot a^2(t) + B \cdot a(t) + D)} = dt, \qquad (10)$$

$$A = -2\Omega_1, \quad B = 2\Omega_3, \quad D = -\left(\Omega_1 + \Omega_3 \cdot \frac{\Omega_3}{\Omega_1}\right).$$

The left part of (10) can be transformed as below:

$$\int \frac{d\,a(t)}{(A \cdot a^2(t) + B \cdot a(t) + D)} = \begin{cases} \frac{2}{\sqrt{\Delta}} \arctan\left(\frac{2A \cdot a(t) + B}{\sqrt{\Delta}}\right), & \Delta > 0 \\ -\frac{2}{\sqrt{-\Delta}} Arth\left(\frac{2A \cdot a(t) + B}{\sqrt{-\Delta}}\right), & \Delta < 0 \end{cases}$$

$$\Delta = (4A \cdot D - B^2) = 8\Omega_1^2 + 4\Omega_3^2 > 0 \;\Rightarrow$$

$$a(t) = \frac{\sqrt{\Delta}}{2A} \tan\left(\frac{\sqrt{\Delta}}{2} t\right) - \frac{B}{2A} \;\Rightarrow\; a(t) = \frac{\Omega_3}{2\Omega_1} - \left(\frac{\sqrt{2\Omega_1^2 + \Omega_3^2}}{2\Omega_1}\right) \tan\left(t \cdot \sqrt{2\Omega_1^2 + \Omega_3^2}\right) \qquad (11)$$

$$b(t) = a(t) - \frac{\Omega_3}{\Omega_1}$$

Thus, we have fully solved system of equations (7) for the partial case (9).

Let us schematically imagine at Fig.1 *e.g.* the appropriate component of velocity field $U$ (depending on time $t$), according to the formulae (6)-(7), which corresponds to the aforementioned *partial* solutions (11) for functions $a(t)$, $b(t)$ via derivation (9)-(11).

We should especially note that variables $\{x,y,z\}$ should be considered as variable parameters in (3)-(4), (6)-(7), (8)-(11) (at Fig.1 we designate $x = t$ just for the aim of presenting the plot of solution).

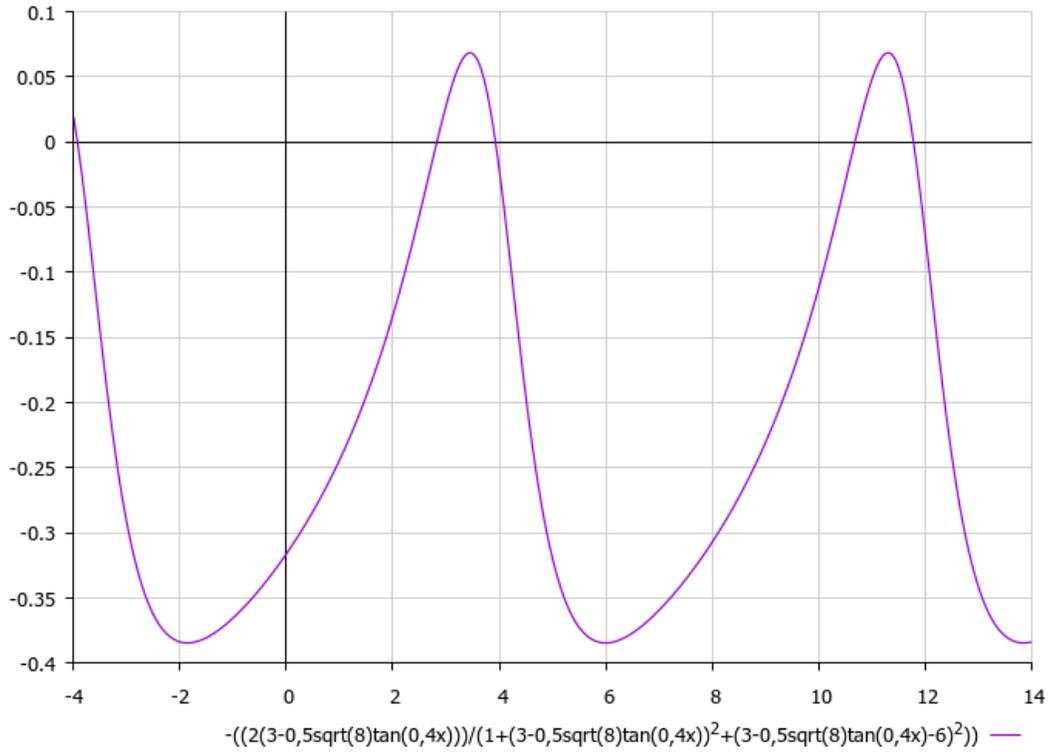

Fig.1. A *schematic* plot of the component $U(t)$ of velocity field (6) for partial solution (11), depending on time $t$

(here we designate $x = t$ just for the aim of presenting the plot of solution)

## 5. Discussion.

As we can see from derivation above, system of Laplace tidal equations (which governs by the dynamics of the ocean that covers the surface of the planet) is proved to be very hard to solve analytically.

Indeed, at first step we should solve the *homogeneous* momentum equation (2) in a form (4) in regard to the time $t$ (depending on 3 components of angular velocity $\vec{\Omega}$ of planetary rotation). In general case, it is almost impossible (even if we would assume each of the aforementioned 3 components of angular velocity equals to the proper constant or to the function, very slowly varying with time *t*); the main reason is that the solution (6) is determined *via* functions *a*, *b*, which are solutions of the mutual system (7) of two *Riccati* ordinary differential equations in regard to the time *t*. Nevertheless, we obtain an elegant way to obtain *the invariant dependence* of function *a*(*t*) on function *b*(*t*) in case $\vec{\Omega} \cong \vec{\Omega}$ (*x*, *y*, *z*) via solving *Jacobi* equation (8) (it should have the proper general solution, see example 1.250 [13]; but meanwhile, such the solution has no simple analytical presentation). So, even at this first step, we could only use the obvious *partial* solution of a simple kind $a = K \cdot b + L$, where coefficients *K*, *L* are determined in conditions (9).

Furthermore, at the second step, we should obtain the *non-homogeneous* part of the solution for velocity field $\vec{v}$ in a form (3) (depending on pressure $p(\vec{r}, t)$, total gravitational potential $\Phi(\vec{r})$, and centrifugal potential $\chi(\vec{r})$ due to planetary rotation). We point out the appropriate algorithm in **Section 2** for obtaining the aforementioned *non-homogeneous* solution for velocity field $\vec{v}$.

The last but not least, at third step we should use the continuity equation in a form (1) or (5) for identifying the spatial components of velocity field $\vec{v}$ (*non-homogeneous* solution) along with the set of functions $\{C_v(x, y, z)\}$ in (3) and $\sigma(x, y, z)$ in (6) (depending on the spatial parts of components of angular velocity $\vec{\Omega}$ (*x*, *y*, *z*) of Earth's rotation, spatial parts of pressure *p*, of total gravitational potential $\Phi$, and centrifugal potential $\chi$).

On the other hand, these fundamental equations (1)-(2) are usually examined with a proper initial and boundary conditions. In this research, we consider only the Cauchy problem in the whole space. Meanwhile, we should note that since the fluid is incompressible for the development above, there is a strong link between initial conditions and the solution inside insofar.

As for the relevance of this new solution (or solving procedure), let us discuss the essential details about the possible physical properties of the aforementioned solution (6)-(8). Equation (8) is known to be the *Riccati* ordinary differential equation, ODE [13]. We also note that due to the special character of the solutions of *Riccati*-type ODEs, there is the possibility for sudden *jumping* in the magnitude of the solution at some time $t_0$ [14-16] (in [15] see e.g. the solutions of *Riccati*-type for *Abel* ordinary differential equation of 1-st order).

In the physical sense, such the aforementioned jumping of *Riccati*-type solutions could be associated with the effect of a sudden acceleration/deceleration of the flow velocity at a definite moment of parametric time $t_0$. This means that there exists a potential for a kind of *gradient catastrophe* [17], depending on the initial conditions.

Ending discussion, we should note that mathematical procedure of obtaining the stationary points (see [18]) for system of Eqns. (2) could be useful and interesting in the sense of exploring a dynamical behavior of tidal waves in vicinity of the aforementioned stationary points or manifolds in ocean basins, governed by Laplace tidal equations (1)-(2).

## 6. Conclusion.

Absolutely new ansatz for solving Laplace tidal equations has been presented in the current research. The aforesaid approach was formulated first in [11] for solving Poisson equations; furthermore, a new type of the solving procedure for Euler-Poisson equations (rigid body rotation over the fixed point) is implemented here for solving momentum equation of Laplace tidal equations also. We have proved here that system of Laplace tidal equations (including continuity equation) could not be solved analytically, taking into account the non-zero potential forces in momentum equation, depending on pressure $p(\vec{r}, t)$, along with total gravitational potential $\Phi(\vec{r})$, and centrifugal potential $\chi(\vec{r})$ (due to planetary rotation). Thus, we fully solved this problem which was formulated 243 years ago by *Pierre-Simon Laplace*.

Meanwhile, momentum equation of LTE is reduced to the system of *Riccati* equations (presenting the time-dependent part of solution), along with the continuity equation (which determines the spatial part of solution); but *Riccati-type* ODE has no analytical solution in general case [13]. Nevertheless, the proper elegant *partial* solution has been obtained here due to invariant dependence between temporary components of the solution.

Also, some remarkable articles should be cited, which concern the problem under consideration, [19]-[27].

## 7. <u>Acknowledgements.</u>

Sergey Ershkov is thankful to Dr. V.V.Sidorenko with respect to review in scientific literature on the tides and tidal interactions as well as in regard to his insightful argumentation during mutual fruitful discussions in the process of preparing of the manuscript.

Authors are thankful to unknown esteemed reviewers as for their having spent the valuable time and efforts as well as in regard to their valuable advices which improved structure of the article significantly.

Remark regarding contributions of authors as below:

In this research, the Roman Shamin (PhD-tutor of Sergey Ershkov) performed the general scientific management of the direction of creative scientific search. Sergey Ershkov is responsible for the results of the article, the obtaining of exact solutions, simple algebra manipulations, calculations, the representation of a general ansatz and calculations of graphical solutions, approximation and also is responsible for the search of partial solutions.

The publication was prepared with the support of the «RUDN University Program 5-100».

## <u>Appendix, A1 (checking of the solution (6)-(7) for equations (4)).</u>

Let us check the solution (6)-(7) for the case $k = 1$ (other cases should be checked in the same manner); such the solution is to be valid for equations (4).

Namely, let us substitute formulae (6) the appropriate functions in (4):

$$U = -\sigma \cdot \left(\frac{2a}{1+(a^2+b^2)}\right), \quad V = -\sigma \cdot \left(\frac{2b}{1+(a^2+b^2)}\right),$$

$$W = \sigma \cdot \left(\frac{1-(a^2+b^2)}{1+(a^2+b^2)}\right),$$

but, in addition to this, accomlishing the aforesaid substituting with the expressions for derivatives (with respect to time) for functions *a* and *b* from (7), where *k = 1*

$$\begin{cases} a' = \dfrac{\Omega_2}{2} \cdot a^2 - (\Omega_1 \cdot b) \cdot a - \dfrac{\Omega_2}{2}(b^2-1) + \Omega_3 \cdot b, \\ \\ b' = -\dfrac{\Omega_1}{2} \cdot b^2 + (\Omega_2 \cdot a) \cdot b + \dfrac{\Omega_1}{2} \cdot (a^2-1) - \Omega_3 \cdot a. \end{cases}$$

Let us begin checking from the first of equations (4) (*k = 1*):

$$\frac{dU}{dt} = \Omega_3 V - \Omega_2 W,$$

whereas other equations should be checked in the same manner. Let's begin:

$$\frac{dU}{dt} = \Omega_3 V - \Omega_2 W, \Rightarrow$$

$$-\sigma \cdot \frac{d}{dt}\left(\frac{2a}{1+(a^2+b^2)}\right) = \left(-\sigma \cdot \left(\frac{2b}{1+(a^2+b^2)}\right)\right) \cdot \Omega_3 - \left(\sigma \cdot \left(\frac{1-(a^2+b^2)}{1+(a^2+b^2)}\right)\right) \cdot \Omega_2, \Rightarrow$$

$$\frac{2\left(\dfrac{da}{dt}\right) \cdot (1+(a^2+b^2)) - 4a\left(a \cdot \dfrac{da}{dt} + b \cdot \dfrac{db}{dt}\right)}{\left(1+(a^2+b^2)\right)^2} = \frac{2b \cdot \Omega_3 + (1-(a^2+b^2)) \cdot \Omega_2}{1+(a^2+b^2)}, \Rightarrow$$

where the proper simlifying of the last equation yields as below:

$$2\left(\frac{\Omega_2}{2}\cdot a^2 - (\Omega_1\cdot b)\cdot a - \frac{\Omega_2}{2}(b^2-1) + \Omega_3\cdot b\right)\cdot(1+(a^2+b^2)) -$$

$$- 4a\left(a\cdot\left(\frac{\Omega_2}{2}\cdot a^2 - (\Omega_1\cdot b)\cdot a - \frac{\Omega_2}{2}(b^2-1) + \Omega_3\cdot b\right) + b\cdot\left(-\frac{\Omega_1}{2}\cdot b^2 + (\Omega_2\cdot a)\cdot b + \frac{\Omega_1}{2}\cdot(a^2-1) - \Omega_3\cdot a\right)\right)$$

$$= (2b\cdot\Omega_3 + (1-(a^2+b^2))\cdot\Omega_2)\cdot\left(1+(a^2+b^2)\right), \Rightarrow$$

$$\Omega_2\cdot a^2 - 2\Omega_1\cdot b\cdot a - \Omega_2\cdot b^2 + \Omega_2 + 2\Omega_3\cdot b +$$
$$+\Omega_2\cdot a^4 - 2\Omega_1\cdot b\cdot a^3 - \Omega_2\cdot b^2\cdot a^2 + \Omega_2\cdot a^2 + 2\Omega_3\cdot b\cdot a^2 + \Omega_2\cdot a^2\cdot b^2 - 2\Omega_1\cdot b^3\cdot a - \Omega_2\cdot b^4 + \Omega_2\cdot b^2 + 2\Omega_3\cdot b^3 -$$
$$-2\Omega_2\cdot a^4 + 4\Omega_1\cdot b\cdot a^3 + 2\Omega_2\cdot b^2\cdot a^2 - 2\Omega_2\cdot a^2 - 4\Omega_3\cdot b\cdot a^2 + 2\Omega_1\cdot a\cdot b^3 - 4\Omega_2\cdot a^2\cdot b^2 - 2\Omega_1\cdot a^3\cdot b + 2\Omega_1\cdot a\cdot b + 4\Omega_3\cdot a^2\cdot b$$
$$= 2b\cdot\Omega_3 + \Omega_2 - a^2\cdot\Omega_2 - b^2\cdot\Omega_2 + 2a^2\cdot b\cdot\Omega_3 + a^2\cdot\Omega_2 - \Omega_2\cdot a^4 - \Omega_2\cdot a^2\cdot b^2 + 2b^3\cdot\Omega_3 + b^2\cdot\Omega_2 - \Omega_2\cdot a^2\cdot b^2 - \Omega_2\cdot b^4, \Rightarrow$$

$$-\Omega_2\cdot a^4 - \Omega_2\cdot b^4 + 2\Omega_3\cdot b^3 - 2\Omega_2\cdot a^2\cdot b^2 + 2\Omega_3\cdot a^2\cdot b + \Omega_2 + 2\Omega_3\cdot b$$
$$= -\Omega_2\cdot a^4 - \Omega_2\cdot b^4 + 2b^3\cdot\Omega_3 - 2\Omega_2\cdot a^2\cdot b^2 + 2a^2\cdot b\cdot\Omega_3 + \Omega_2 + 2b\cdot\Omega_3 ,$$

so, we have obtained the valid equality (it means that the checking of the first of equations (4) is successfully finished).

Let us perform also the checking of the second of equations (4) in the similar way:

$$\frac{dV}{dt} = \Omega_1 W - \Omega_3 U, \Rightarrow$$

$$-\sigma\cdot\frac{d}{dt}\left(\frac{2b}{1+(a^2+b^2)}\right) = \left(\sigma\cdot\left(\frac{1-(a^2+b^2)}{1+(a^2+b^2)}\right)\right)\cdot\Omega_1 + \left(\sigma\cdot\left(\frac{2a}{1+(a^2+b^2)}\right)\right)\cdot\Omega_3, \Rightarrow$$

$$-\frac{\left(2\left(\dfrac{db}{dt}\right)\cdot(1+(a^2+b^2)) - 4b\left(a\cdot\dfrac{da}{dt}+b\cdot\dfrac{db}{dt}\right)\right)}{\left(1+(a^2+b^2)\right)^2} = \frac{2a\cdot\Omega_3 + (1-(a^2+b^2))\cdot\Omega_1}{1+(a^2+b^2)}, \Rightarrow$$

$$2\left(-\frac{\Omega_1}{2}\cdot b^2 + (\Omega_2 \cdot a)\cdot b + \frac{\Omega_1}{2}\cdot(a^2-1) - \Omega_3\cdot a\right)\cdot(1+(a^2+b^2)) -$$

$$- 4b\left(a\cdot\left(\frac{\Omega_2}{2}\cdot a^2 - (\Omega_1\cdot b)\cdot a - \frac{\Omega_2}{2}(b^2-1) + \Omega_3\cdot b\right) + b\cdot\left(-\frac{\Omega_1}{2}\cdot b^2 + (\Omega_2\cdot a)\cdot b + \frac{\Omega_1}{2}\cdot(a^2-1) - \Omega_3\cdot a\right)\right)$$

$$= -\left(2a\cdot\Omega_3 + (1-(a^2+b^2))\cdot\Omega_1\right)\cdot\left(1+(a^2+b^2)\right),$$

where the proper simlifying of the last equation yields as below:

$$-\Omega_1\cdot b^2 + 2\Omega_2\cdot a\cdot b + \Omega_1\cdot a^2 - \Omega_1 - 2\Omega_3\cdot a - \Omega_1\cdot a^2\cdot b^2 + 2\Omega_2\cdot a^3\cdot b + \Omega_1 a^4 - \Omega_1 a^2 - 2\Omega_3\cdot a^3 +$$

$$-\Omega_1\cdot b^4 + 2\Omega_2\cdot a\cdot b^3 + \Omega_1\cdot b^2\cdot a^2 - \Omega_1\cdot b^2 - 2\Omega_3\cdot b^2\cdot a -$$

$$-2\Omega_2\cdot a^3 b + 4\Omega_1\cdot b^2\cdot a^2 + 2\Omega_2 ab^3 - 2\Omega_2 ab - 4\Omega_3\cdot ab^2 + 2\Omega_1\cdot b^4 - 4\Omega_2\cdot a\cdot b^3 - 2\Omega_1\cdot a^2 b^2 + 2\Omega_1\cdot b^2 + 4\Omega_3\cdot ab^2$$

$$= -2a\cdot\Omega_3 - \Omega_1 + a^2\cdot\Omega_1 + b^2\cdot\Omega_1 - 2a^3\cdot\Omega_3 - a^2\Omega_1 + a^4\Omega_1 + b^2\cdot a^2\Omega_1 - 2ab^2\cdot\Omega_3 - b^2\Omega_1 + a^2 b^2\Omega_1 + b^4\Omega_1, \Rightarrow$$

$$\Omega_1\cdot a^4 + \Omega_1\cdot b^4 + 2\Omega_1\cdot(b^2\cdot a^2) - 2\Omega_3\cdot a^3 - 2\Omega_3\cdot(b^2\cdot a) - 2\Omega_3\cdot a - \Omega_1$$

$$= a^4\cdot\Omega_1 + b^4\cdot\Omega_1 + 2(a^2\cdot b^2)\cdot\Omega_1 - 2a^3\cdot\Omega_3 - 2(a\cdot b^2)\cdot\Omega_3 - 2a\cdot\Omega_3 - \Omega_1 ,$$

so, we have obtained the valid equality again (it means that the checking of the second of equations (4) is successfully finished).

At last but not least, let us perform also the checking of the third of equations (4) in the similar way as above:

$$\frac{dW}{dt} = \Omega_2 U - \Omega_1 V, \Rightarrow$$

$$\sigma \cdot \frac{d}{dt}\left(\frac{1-(a^2+b^2)}{1+(a^2+b^2)}\right) = \left(-\sigma \cdot \left(\frac{2a}{1+(a^2+b^2)}\right)\right) \cdot \Omega_2 + \left(\sigma \cdot \left(\frac{2b}{1+(a^2+b^2)}\right)\right) \cdot \Omega_1, \Rightarrow$$

$$\frac{\left(-2\left(a \cdot \frac{da}{dt} + b \cdot \frac{db}{dt}\right) \cdot (1+(a^2+b^2)) - 2(1-(a^2+b^2)) \cdot \left(a \cdot \frac{da}{dt} + b \cdot \frac{db}{dt}\right)\right)}{\left(1+(a^2+b^2)\right)^2} = \frac{2b \cdot \Omega_1 - 2a \cdot \Omega_2}{1+(a^2+b^2)}, \Rightarrow$$

$$-4 \cdot \left(a \cdot \frac{da}{dt} + b \cdot \frac{db}{dt}\right) = \left(2b \cdot \Omega_1 - 2a \cdot \Omega_2\right) \cdot \left(1+(a^2+b^2)\right),$$

where the proper simlifying of the last equation yields as below:

$$-4 \cdot \left(a \cdot \frac{da}{dt} + b \cdot \frac{db}{dt}\right) = \left(2b \cdot \Omega_1 - 2a \cdot \Omega_2\right) \cdot \left(1+(a^2+b^2)\right)$$

$$2\Omega_2 \cdot a^3 - 4\Omega_1 \cdot b \cdot a^2 - 2\Omega_2 ab^2 + 2\Omega_2 a + 4\Omega_3 \cdot ab - 2\Omega_1 \cdot b^3 + 4\Omega_2 \cdot a \cdot b^2 + 2\Omega_1 \cdot ba^2 - 2\Omega_1 \cdot b - 4\Omega_3 \cdot ab =$$

$$= -2b \cdot \Omega_1 + 2a \cdot \Omega_2 - 2a^2 \cdot b \cdot \Omega_1 + 2a^3 \cdot \Omega_2 - 2b^3 \cdot \Omega_1 + 2a \cdot b^2 \cdot \Omega_2, \Rightarrow$$

$$2\Omega_2 \cdot a^3 - 2\Omega_1 \cdot b^3 + 2\Omega_2 \cdot (a \cdot b^2) - 2\Omega_1 \cdot (b \cdot a^2) + 2\Omega_2 \cdot a - 2\Omega_1 \cdot b =$$

$$= 2a^3 \cdot \Omega_2 - 2b^3 \cdot \Omega_1 + 2(a \cdot b^2) \cdot \Omega_2 - 2(a^2 \cdot b) \cdot \Omega_1 + 2a \cdot \Omega_2 - 2b \cdot \Omega_1,$$

so, we have obtained the valid equality once again (it means that the checking of the third of equations (4) is also successfully finished).

**Appendix, A2 (solving procedure for equation (8)).**

Equation (8) has obvious partial solution $a = K \cdot b + L$ ($K = const$, $L = const$) for which are valid the proper conditions below:

$$\begin{cases} -\frac{1}{2}\Omega_1 \cdot K(K+1) - \frac{1}{2}\Omega_2 \cdot (K^2+1) = 0, \\ -\Omega_1 \cdot L(K+1) + \Omega_3 \cdot (K+1) = 0, \\ -\frac{1}{2}\Omega_1 \cdot (L^2-1) + \frac{1}{2}\Omega_2 \cdot (L^2+1) + \Omega_3 \cdot L = 0, \end{cases} \Rightarrow \begin{cases} \Omega_2 = -\Omega_1 \cdot \frac{K(K+1)}{(K^2+1)}, \\ (K+1) \cdot \left(\Omega_3 - \Omega_1 \cdot L\right) = 0, \\ \Omega_1 \cdot (L^2+1) \cdot \left(\frac{K-1}{(K^2+1)}\right) = 0, \end{cases} \Rightarrow \begin{cases} \Omega_2 = -\Omega_1, \\ L = \frac{\Omega_3}{\Omega_1}, \\ K-1 = 0. \end{cases} \quad (9)$$

Let us check the solution; taking into account (9), we should substitute $a = K \cdot b + L$ into (8):

$$K\frac{\Omega_1}{2} \cdot b^2 - (\Omega_2 \cdot (K^2 b + KL)) \cdot b - \frac{\Omega_1}{2} \cdot (K^2 b^2 + 2KLb + L^2 - 1) + \Omega_3 \cdot (Kb + L) =$$

$$= -\frac{\Omega_2}{2} \cdot (K^2 b^2 + 2KLb + L^2) + (\Omega_1 \cdot b) \cdot (Kb + L) + \frac{\Omega_2}{2}(b^2 - 1) - \Omega_3 \cdot b, \Rightarrow$$

$$\left(-\frac{1}{2}\Omega_1 \cdot K(K+1) - \frac{1}{2}\Omega_2 \cdot (K^2+1)\right) \cdot b^2 + \left(-\Omega_1 \cdot L(K+1) + \Omega_3 \cdot (K+1)\right) \cdot b +$$

$$+\left(-\frac{1}{2}\Omega_1 \cdot (L^2-1) + \frac{1}{2}\Omega_2 \cdot (L^2+1) + \Omega_3 \cdot L\right) = 0, \Rightarrow$$

$$\Omega_1^2 \cdot b^2 + 2\Omega_1 \cdot \Omega_3 \cdot b + \left(\frac{\Omega_3^2}{\Omega_1}\right) \cdot \Omega_1 = (b \cdot \Omega_1 + \Omega_3)^2$$

- it means, indeed, that equality $a = K \cdot b + L$, along with demand (9), is the partial case of solution of equation (8).